# Specular Reflection from the Great Pyramid at Giza

Donald E. Jennings
*Goddard Space Flight Center, Greenbelt, Maryland, USA (retired)*
*email: seked2500@gmail.com*

## Abstract

The pyramids of ancient Egypt are said to have shone brilliantly in the sun. Surfaces of polished limestone would not only have reflected diffusely in all directions, but would also likely have produced specular reflections in particular directions. Reflections toward points on the horizon would have been visible from large distances. On a particular day and time when the sun was properly situated, an observer stationed at a distant site would have seen a momentary flash as the sun's reflection moved across the face of the pyramid. The positions of the sun that are reflected to the horizon are confined to narrow arcs in the sky, one arc for each side of the pyramid. We model specular reflections from the pyramid of Khufu and derive the annual dates and times when they would have been visible at important ancient sites. Certain of these events might have coincided with significant dates on the Egyptian calendar, as well as with solar equinoxes, solstices and cross-quarter days. The celebration of Wepet-Renpet, which at the time of the pyramid's construction occurred near the spring cross-quarter day, would have been marked by a specular sweep of sites on the southern horizon. On the autumn and winter cross-quarter days reflections would have been directed to Heliopolis. We suggest that on those days the pyramidion of Khafre might have been visible in specular reflection over the truncated top of Khufu's pyramid.

[*Subjects*: Egyptian history, pyramids, optics, solar astrometry]

**Introduction**

The great pyramids of Egypt at Giza were once encased in polished white limestone that shone brightly in reflected sunlight [1]. Descriptions of this phenomenon seem to refer to diffuse reflection, an efficient scattering of light in all directions. Depictions of the pyramids often show their sides glowing in sunlight, with little dependence on the position of the sun in the sky. But the skill of the ancient Egyptians in smoothing and buffing stone would likely have produced, in addition to the diffuse aspect, a specular component to the reflection. This mirror-like effect would have sent a portion of the light in a particular direction, a direction that would have changed as the sun moved in the sky. If observed from a distance at just the right location and time, a specular reflection would have caused a momentary brightening of the pyramid. Egyptians would have been familiar with specular reflection from, for instance, the sun's glint off a surface of calm water. In this paper we model the geometry of specular reflections from the Great Pyramid of Khufu and examine how they might have been directed to nearby cultural sites.

Imagine standing at Heliopolis in early February 2560 BCE looking southwest across the Nile valley. The recently completed giant pyramid, called the "horizon of Khufu", gleams white and dominates the horizon. Suddenly, just before noon, the pyramid begins to brighten and is soon shining like a second sun, mimicking a sunset. The glare lasts some moments before dimming. For that brief time the great pyramid has controlled the sun, sending a piece of it to its home, the temple of Ra, where you stand. Although you probably know in advance that this event will occur, it nevertheless leaves you astonished. It is a clear demonstration of the connection between your king and the sun god.

Although the author knows of no ancient record of these displays, if the surfaces of the pyramid were indeed polished the spectacle surely took place. We know that the Egyptians were capable of applying a quality of polish to stone that would support specular reflection. The high-quality Tura limestone used for the pyramid's outer casings can be worked to such a shine. Limestone has a hardness like that of marble and can take a comparable polish [2]. Given its refractive index of 1.6 [3], Fresnel reflection from a polished limestone surface sends 5% of the incident light in the specular direction. This is similar to that from a glass window. Modern glass or marble structures, including pyramids, exhibit specular reflections. When viewed from the proper direction the specular component is many times brighter than the diffuse portion of reflected sunlight. At the time of completion, the original pyramid casings may have reflected close to the full 5%, but even a lower-grade polish or one deteriorating over time would have produced a noticeable reflection. A slightly rough, uneven surface would have reduced the brightness and spread out the beam, but the display would still have been remarkable. The 0.5° width of the solar disk allows some surface imperfection without greatly degrading the sun's image. Moreover, an aged surface might be temporarily rejuvenated if wetted by even a light rain. If a portion of the pyramid (or its pyramidion capstone) were gold-covered, the reflectivity would have been more like 50% and the specular intensity from that part would have been ten times greater than from the bare limestone. Our analysis applies just as well to the gilded portion of the pyramid.

Strong specular reflections can still be seen on two recovered capstones, the granite pyramidions of Amenemhet III (1820 BCE) and Khendjer (1760 BCE) [4], when they are illuminated by a directed light source. Those pyramidions, as well as others, are intricately engraved with inscriptions, including cartouches of the pharaohs. When the sun is reflected toward its target any engravings on the reflecting surface are, in a sense, imprinted on the outgoing beam. In this way the son of Ra could pass his identity to a distant shrine.

Although ten's of kilometers away, many early cultural sites were in direct line of sight of the Giza pyramids. As seen from Giza, these locations would be on the horizon. A solar flash at one of these sites would have lasted several minutes as the sun moved in the sky and its image moved across the pyramid. At ranges greater than 12 km the angular width of the pyramid of Khufu is smaller than the 0.5° angular diameter of the sun, so the sun's width would determine the duration of the event. At distances beyond about 43 km a site would be below the horizon and out of sight of the Great Pyramid (rare atmospheric conditions might extend the range). In any case, a beam reflected in the direction of any site would be a gesture of acknowledgement.

**Modeling the Reflections**

We begin by calculating the locations of the sun corresponding to specular reflections toward observation points on the horizon (see Figure 1). The relationship between the incident solar position and the direction of the outgoing reflected beam depends on the orientation of the pyramid and the slope angle of its sides. Khufu's pyramid, like all of the major pyramids, is aligned to the cardinal directions (north-south, east-west) and its sides are slanted at 51.8 degrees (the angle of the normal with respect to vertical) [5]. Confining the reflected beam to the horizontal plane restricts the sun to a narrow range in the sky. Each face of the pyramid has its own arc of permissible solar coordinates that project to the horizon. We model each arc by calculating the sun's local azimuth and elevation as a function of the azimuth of the horizontal reflected beam. Appendix 1 describes how the equations for the solar coordinates are derived.

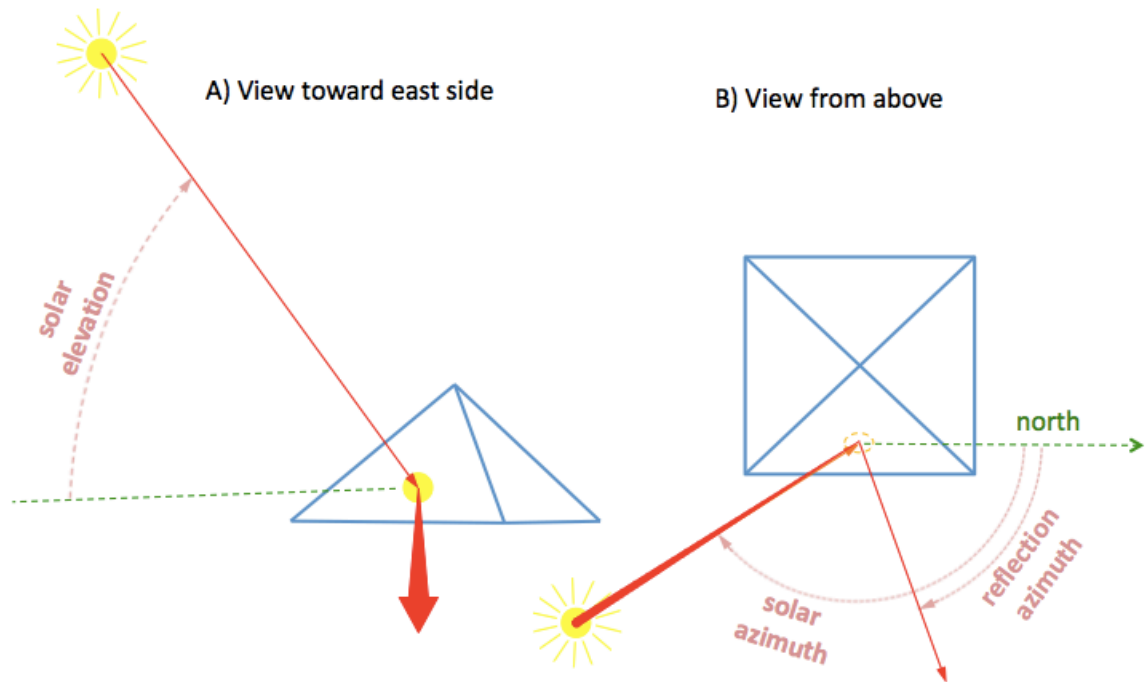

**Figure 1**. When the sun was in the proper sky position a distant observer would see a reflection from the side of the pyramid. A) Reflection from the east face toward the observer. B) Top view of the same reflection geometry. For a given observer location the required solar position is fixed by the orientation and slope of the pyramid. The direction of the sun is measured in azimuth from due north and in elevation above the horizontal plane. The azimuthal direction of the observer in the horizontal plane is measured from due north.

Four arcs of solar position, one for each side of the pyramid, are plotted in Figure 2. Only a limited portion of each arc is actually accessible because the sun must lie between the extremes of winter and summer solstices. These limits are set by the latitude of the Giza necropolis (30 N) and the Earth's obliquity (23.4°, see below). From Figure 2 it is apparent that almost every point on the horizon was accessible to a reflection from the pyramids at some time during the year. Most azimuths from 38° to 322° were reached by at least one of the four sides of the Khufu pyramid. There were exceptions, however. Of the four cardinal points, only the due-south horizon (180° azimuth) could be illuminated. Due east and due west were approached, but not reached, on the summer solstice. No reflection came close to due north. Also, once it was completed in 2532 BCE, a swath to the southwest was blocked by the pyramid of Khafre. The symmetry of the arcs about 180° solar azimuth in Figure 2 is a consequence of the pyramid being aligned to north.

**Selected Sites and Events**

With this model we can find the annual dates and times of specular events at sites in the vicinity of Giza and attempt to match them with prominent dates on the Eqyptian calendar. Table 1 lists examples of places toward which reflections from the Great Pyramid would have been directed. The sites in the table either existed at the time of Khufu (2566 BCE) or were established soon after. Every solar position (except on the solstices) occurs on two dates, one during winter-spring and the other during summer-autumn. In the table, the reflection azimuth and distance were found using the Movable Type Scripts online calculator [6]. The sun azimuth and elevation for each site was then derived using our model (Appendix 1). From these solar coordinates the date and time for each event was determined using the online Solar Calculator of the National Oceanic and Atmospheric

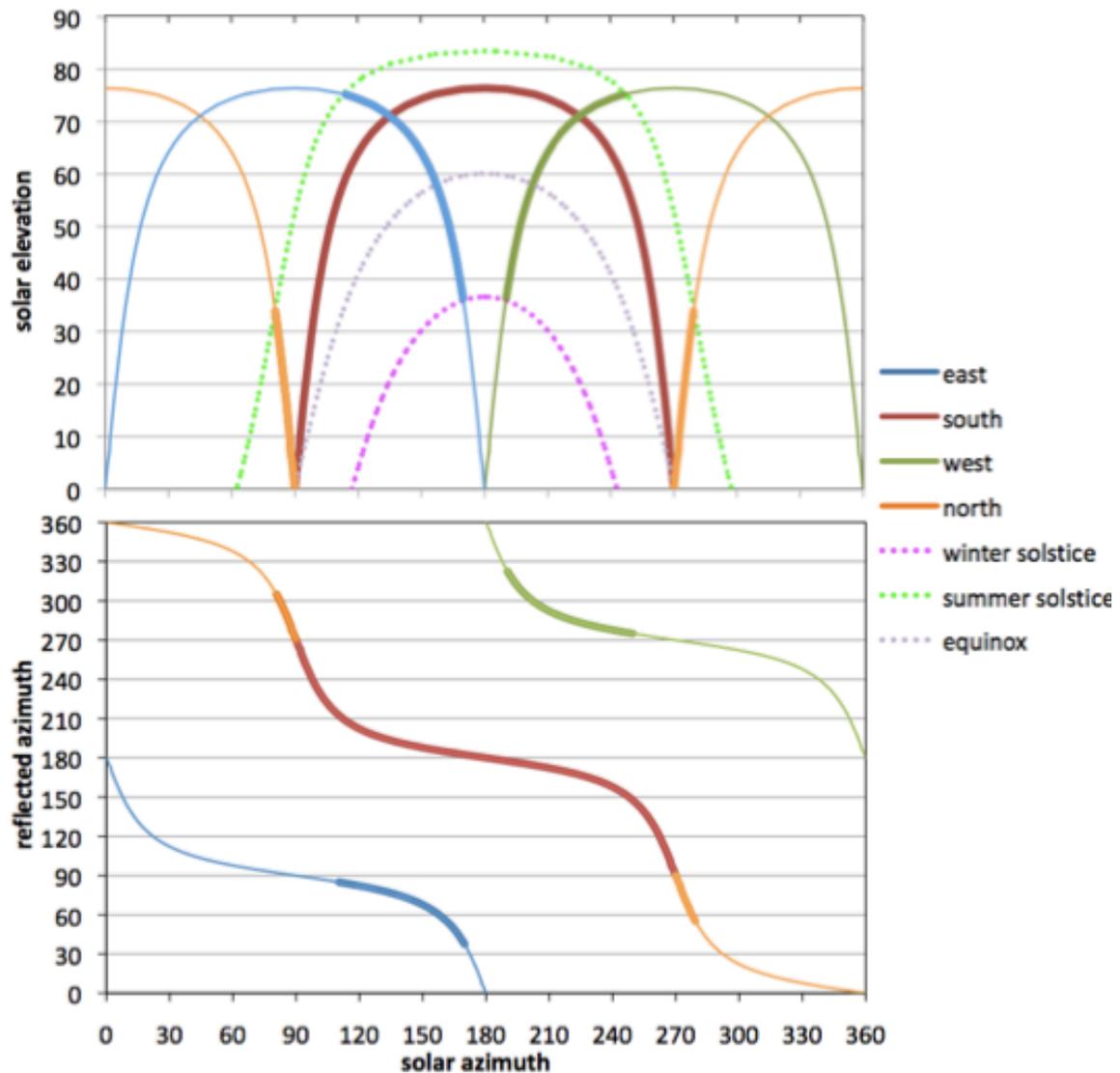

**Figure 2.** Sun positions that produced horizontal reflections from the faces of the Khufu pyramid. A curve of solar position is shown for each of the four sides (thin solid lines). Reachable solar locations (bold solid lines) fall between the limits of winter and summer solstices (dotted lines). To use the chart: 1) begin with the desired reflected azimuth and use the lower plot to find the solar azimuth; 2) from the solar azimuth use the upper plot to find the solar elevation. These solar coordinates can then used to find the date and time of the event. Elevation is measured from horizontal and azimuth is measured from north (see Figure 1).

Administration (NOAA) [7]. This calculation uses the current obliquity of the Earth, 23.4°. The ~0.5° larger obliquity in 2500 BCE would shift our April/May and August event dates closer to equinox by about one day. Since this difference would not materially affect our conclusions, the adjustments have not been applied in Table 1. Estimated overall accuracy of our calculated dates is ±2 days.

Table 1. Specular Reflection Events from the Pyramid of Khufu

| Site | Reflection* | | Sun | | Winter-Spring Date** | Summer-Autumn Date** | Local solar time hh:mm | Dist-ance* (km) | Age (BCE) |
| --- | --- | --- | --- | --- | --- | --- | --- | --- | --- |
| | Az (°) | Side | Az | El | | | | | |
| Heliopolis[†] | 44.9 | East | 166.8 | 43.3 | Feb 6 | Nov 5 | 11:20 | 23.6 | 2700 |
| Djedefre Pyramid | -44.1 | West | 192.8 | 42.6 | Feb 4 | Nov 7 | 12:39 | 8.2 | 2558 |
| Memphis[†] | 141.0 | South | 253.8 | 49.1 | Apr 23 | Aug 18 | 14:40 | 18.7 | 3200 |
| Userkaf Sun Temple | 144.4 | South | 251.8 | 52.2 | Apr 25 | Aug 17 | 14:26 | 10.8 | 2458 |
| Djoser Step Pyramid | 146.4 | South | 250.5 | 54.1 | Apr 26 | Aug 16 | 14:18 | 14.4 | 2630 |
| Sneferu Red Pyramid | 159.9 | South | 237.2 | 65.9 | May 1 | Aug 10 | 13:23 | 20.2 | 2590 |
| Amenemhet I Pyramid[††] | 168.9 | South | 219.9 | 72.5 | May 4 | Aug 8 | 12:47 | 45.8 | 1950 |
| Meidum Pyramid[††] | 178.1 | South | 188.2 | 76.3 | May 5 | Aug 6 | 12:08 | 65.8 | 2589 |
| Khufu Causeway | 76.5 | East | 135.6 | 70.9 | May 2 | Aug 9 | 11:05 | horizon | 2566 |
| Due South | 180.0 | South | 180 | 76.4 | May 5 | Aug 6 | 12:00 | | |
| Solstice Noon[§] | 180.0 | South | 180 | 83.5 | Jun 20 | Jun 20 | 12:00 | 0.99 | |

*Directions and distances were found using the Movable Type Scripts calculator [6].
**Dates and times were found using the NOAA Solar Calculator [7]. Dates are Gregorian.
[†]Location of Heliopolis: obelisk. Location of Memphis: temple of Ptah.
[††]Amenemhet I and Meidum sites are too distant for Giza to have been directly visible.
[§]Viewing the apex. Distance south from edge of pyramid, 989 m. North, 296 m.

*Heliopolis*

Chief among the sites of ancient importance was Heliopolis and it is well known that the pyramids of Giza have an intimate geometric relationship with the religious center. The southeastern corners of the three principal pyramids at Giza are aligned on a 45° diagonal with Heliopolis. The apexes of the two largest, those of Khufu and Khafre, are similarly aligned with Heliopolis (along with the northwest corner of Menkaure's). Today, the one remnant of ancient Heliopolis, the obelisk of the Temple of Ra-Atum, is precisely on the 45° bearing from the Giza tombs. Although it is not visible today, the city once had a clear view of the Giza Plateau. Heliopolis was believed to be the creation site and home of the sun god Ra, with whom the kings identified. They would have sought phenomena that tied their tombs to the sacred city.

The Giza necropolis is located at the unique spot where the 30 N latitude line crosses the 45° line from Heliopolis. At 30 N the celestial north pole is one-third of the way to the local zenith, while the 45° diagonal is halfway between north and east. This conjunction of simple fractions would have counted favorably with the builders. Additionally, the Great Pyramid is due south of ancient Letopolis (modern Ausim), another key religious city. The Giza site had the practical advantage of being located on the visually prominent raised edge of the western desert where the limestone bedrock was particularly favorable for supporting massive structures. The location must have seemed propitious to the pyramids' architects.

A solar flash at Heliopolis came from the east side of the Khufu pyramid. It was brightest on February 6 and November 5, both at 40 minutes before noon (Table 1). At a distance of 24 km the duration of the event would have been limited to a few minutes by the sun's angular width. February 6 is close to the winter cross-quarter day, halfway between winter solstice and spring equinox. That date is close to the mid-point of the period of emergence and growth as the Nile receded in its yearly cycle. November 5 is close to the autumnal cross-quarter day between autumnal equinox and winter solstice, and is near the end of the Nile inundation. Cross-quarter days are significant because, together with the quarter days (solstices and equinoxes), they mark the eight cardinal points of the solar year. They would have provided a solar reference for tracking the Nile flood cycle. The Nile cycle was naturally tied to the solar cycle, which by the time of Khufu was noticeably out-of-sync with the Egyptian civil calendar [8]. The other two cross-quarter days, in spring and summer, were marked by sweeps of the sun's reflection along the southern horizon and stood in proximity to the setting and rising of Sirius (see below).

There is a curious circumstance in the alignment of the Giza pyramids with Heliopolis. By placing the southeast corner of the three pyramids on the 45-degree diagonal, the pyramids of Khafre and Menkaure are hidden behind that of Khufu when viewed from Heliopolis. This must have been a major compromise for pharaoh Khafre, the second to build at the site. The obscuration meant that the sun's reflection from Khafre's pyramid would be blocked from projection to Heliopolis. But Khafre may have found a solution to this problem. The top 9 meters of the Khufu pyramid is missing today [5] and it is not known whether it was ever present. The height of Khafre's pyramid plus the higher elevation of its base combine to place its summit near the same height as the virtual tip of the Khufu pyramid. If the top of Khufu's pyramid were missing at the time that Khafre completed his construction, several meters of the pyramidion of Khafre might have been visible from Heliopolis over Khufu's truncated summit. Khafre's pinnacle would have seemed to form the tip of Khufu's pyramid. Twice a year its specular flashes would have been seen at Heliopolis. Assuming that the slope of the sides of Khafre's pyramidion was the same as the rest of the pyramid, the flash would have taken place about eight minutes earlier than the flash from Khufu's pyramid. If the pyramidion of Khafre were gold or gold-coated, the reflection would have been especially bright. Did Khafre intentionally leave off or remove the top of his father's pyramid to gain a view from Heliopolis? Even without intense specular reflections a flattened top would have permitted a line-of-sight connection. Perhaps, as both a tribute and a bit of upstaging, Khafre found a way to link his pyramid to the temple of the sun and to share the sacred bond with his father. From the available survey information it appears plausible that the pinnacle of Khafre could be seen from Heliopolis over Khufu's flattened top. A more detailed comparison of heights, plus a more complete history of Khufu's capstone, is needed to decide whether this conjecture is tenable.

*Djedefre Pyramid*

The pharaoh Djedefre was the first son of Khufu to succeed him. Djedefre is perhaps the first to have deemed the sun god Ra to be above all other gods and to identify himself as the son of Ra [9]. His pyramid at Abu Roash, 8 km north of Giza, was next to be built after his father's great pyramid. Looking north from the pyramid of Khufu, Djedefre's pyramid stands at -44.1°, to the west, in near-symmetry with Heliopolis 44.9° to the east. (It is interesting that the northeast corners of the Khufu and Djedefre pyramids are along a -44.7° line, closely mirroring the +45° alignment of the southeast corners of the Khufu, Khafre and Menkaure pyramids.) As a result, for each reflection seen at Heliopolis from the east side of Khufu's pyramid there was a counterpart seen at Djedefre's pyramid from the west side. Because Djedefre's bearing was not precisely -45°, the peak flashes at the two sites occurred two days apart, in early February and again in early November, 40 minutes before noon at Heliopolis and 39 minutes after noon at the pyramid of Djedefre (Table 1). Moreover, Djedefre's pyramid, if polished, would have reflected toward Giza. A specular flash would have been sent from its south side toward Khufu's pyramid on April 19 and again on August 22 (the pyramid had a ~52° slope, similar to Khufu's [10]). Heliopolis and the pyramid of Djedefre fall on a 64° diagonal that points west near sunset on the winter solstice and east near sunrise on the summer solstice [11]. Because of this alignment, the pyramid of Djedefre reflected an image of the sun eastward toward Heliopolis on the equinoxes (it is not known whether a second pyramid at Abu Roash, Lepsius 1, would have blocked the beam). Where this diagonal crosses the -45° line from Khufu's pyramid may have fixed the location of Djedefre's tomb. (With a compromise of 0.9° from 45° the builders may have taken advantage of a natural mound at Abu Roash.) Djedefre's position also places Letopolis on a 31° azimuth, halfway between due north and Heliopolis.

*Southern Sweep*

During the two weeks leading up to May 5, specular beams were sent toward principal sites on the southern horizon. The illuminations took place on separate days in succession beginning in the southeast with Memphis on April 23 and reaching due south on May 5 (Table 1). Likewise, on August 6 and the two weeks following, the solar reflections moved along the horizon from due south to Memphis. During these periods in mid-spring and mid-summer the "south" arc in Figure 2 is nearly coincident with the midday path of the sun, so that a large section of the southern horizon received the specular flashes over a relatively few days. Because major sites were located near the Nile to the southeast, each flash event took place in the afternoon. Dates of the southern transit were determined by the 51.8° slope of the pyramid's sides. A horizontal reflection was directed due south (180° azimuth) when the noon sun was at 76.4° elevation, its position on May 5 and August 6. Together with the February 6 and November 5 flashes toward Heliopolis, these dates comprise the four cross-quarter days of the year. The May and August dates coincide roughly with the heliacal setting and rising of Sirius, whose annual return marked the start of the Nile flood and the beginning of the Egyptian yearly agricultural cycle. In the time of Khufu the civil New Year festival Wepet-Renpet would have been celebrated in early May [8].

Some southern pyramids and temples seem to have been purposefully arranged with respect to Heliopolis [11]. Khufu's pyramid may have supplemented this arrangement by providing a secondary solar reference point, forming a three-way connection in each case. The Great Pyramid would send rays of the sun to Heliopolis in February, and then in April/May to each of the southern sites. This would be important at Abu Sir, Saqqara and Dashur where the line of sight to Heliopolis was obstructed by an intervening ridge while the view of Giza was clear. Other sites further north, like Abu Gorab, were in view of both Heliopolis and Giza.

*The Causeway*

The New Year festival Wepet-Renpet is tied even more closely to Khufu when the causeway to his tomb is considered. The causeway connected his mortuary temple at the center of the east side of the pyramid with his valley temple at the edge of the Nile. Taking the angle of the causeway as 13.5° north of east [12] its bearing bisects the angle to sunrise on the summer solstice, 28° north of east. The causeway thus marks the geometric midpoint between the equinox and the solstice. Belmonte [8] suggests that the causeway points to the sunrise on Wepet Renpet circa 2550 BCE. Sunrise at 13.5° occurred on April 19, sixteen days before the temporal cross-quarter day, May 5, but perhaps within the period of the New Year festivities. On May 2, an hour before noon, the pyramid produced its own conjugate "sunrise" by sending a reflection from its east side along the causeway toward 13.5° on the horizon (Table 1). Both the real and reflected sunrises, plus the sweep of the southern horizon, might have been part of a celebration that took place during the span of days associated with Wepet-Renpet and culminating on May 5 with the projection of the sun to the southern-most horizon.

*Solstices and Equinoxes*

As described above, the four cross-quarter days, halfway between the equinoxes and solstices, are associated with noteworthy specular events. Are there also exceptional alignments that coincide with the solstices and equinoxes? As can be seen in Figure 2, there are two times on the winter solstice and four times on the summer solstice when the sun's reflection was sent to the horizon. The two on winter solstice represent the farthest north that the reflected beam ever reached (38° and 322° azimuth). On summer solstice two of the reflections, toward the east and west, were from the north face and were at the steepest of all incidence angles from that face. They were therefore the brightest reflections from that face as seen from a distance. The other two instances on summer solstice were from the east and west sides. In those cases the reflected beam came the closest of all reflections to due east and due west (84° and 276° azimuth). The due north (0° azimuth) horizon could not be reached from any side and, in particular, no reflection could reach Letopolis.

At noon on the summer solstice the sun was too high in the sky to cast a horizontal reflection. Its elevation at noon on June 20 is 83.5° elevation, above the 76.4° needed to put the reflected beam at the horizon (that geometry occurred on May 5). Instead, the beam would have been angled downward, 7° below the horizon, toward the ground south of the pyramid. The reflection could have been viewed at distances out to about 1 km due south. At the same time the north face was illuminated at a steep incidence angle that caused the reflection to be cast downward 20° below the horizon toward the north. The northward reflection could be viewed within about 0.3 km of the base of the pyramid.

An intriguing reflection geometry took place on the winter solstice. At noon on December 20, when the sun was at 36.6° elevation, the sun's reflection from the south face of the Khufu pyramid was at 39.8°, almost directly back on itself. In this arrangement the sun in a sense was "seeing" its own reflection. This peculiar alignment was a result of the 51.8° slope of the pyramid's sides, although it is not likely that the angle was chosen to produce the effect. In fact, the slope of the pyramid is too small to have produced a perfect back-reflection, hence the 3.2° angle between the incident and reflected beams. The difference would have been even greater in 2500 BCE, since the obliquity of the Earth was larger at that time. Khafre's pyramid, with its steeper slope, would have produced a more precise back reflection.

On the days of the equinoxes there were two reflections toward the horizon, at ±61° azimuth, from the east and west faces. These were close to the directions of sunrise and sunset on the summer solstice (±62°). Thus the sun that rose due east on the equinox was cast, just before noon, to where the sun rises on the solstice. A rule of thirds is apparent here: the 45° angle between equinox sunrise and Heliopolis is sectioned into three roughly equal parts by the causeway (14°) and solstice sunrise (28°). The Egyptians likely noted these simple geometric ratios, and the associated specular events may have enhanced the relationship. The analogous reflections toward the winter solstice sunrise and sunset points (±117° azimuth) took place in the afternoon and morning, respectively, on both April 8 and September 3. It is not known if these dates would have held any significance.

**Discussion**

It is fascinating to imagine what the great pyramids originally looked like under the bright Egyptian sky. Thinking about their appearance in reflected sunlight when viewed from afar may add a new aspect to their function. But whether the great pyramids produced specular reflections cannot be known for sure. Without evidence from the archeological record we cannot be certain that the limestone casing was polished to a mirror finish. We can nevertheless be sure that reflections from calm water, polished granite and metal mirrors were familiar in 2500 BCE. The Egyptians would have understood how a mirror could be used to cast a sunbeam in any direction. (Magli [11] has suggested that a mirror was installed at Heliopolis to send signals to various sites, including Giza.) We also know that the pyramidions capping some pyramids were highly polished granite [4] and we believe that the top portions of some pyramids were gold coated. Either of these would have produced reflections of the type we have modeled here, with those from gold being especially intense. We therefore feel justified in investigating how the sun would have interacted with the pyramid if it originally had a specular finish. If it did produce specular reflections, beams would surely have been sent toward the sites and directions listed in Table 1 and discussed above. Without historical support, however, we cannot know whether the Egyptians held all or any of those events in importance.

An objection might be raised that the identified matchups of specular events with notable sites and dates are too loose or uncertain to permit serious conclusions to be drawn. The importance of the dates of the events and the relevance of seeing the solar flashes from distant sites may be overestimated. However, the timing of the events is not arbitrary. A small change in the slope angle has a large effect. The slightly steeper sides of Khafre' Pyramid shift the southern sweep by nine days from the dates for Khufu, earlier in spring and later in summer. (Interestingly, the Heliopolis dates in February and November do not change.) Also, pyramids located near 30 N latitude with slopes less than 48° cannot reflect the sun toward the due-south horizon. The slope of Khufu is in a small range between being too shallow to access the southern horizon and too steep to be stable.

We have concentrated on views from a distance but, as with the summer solstice event described above, specular reflections could also have been seen from locations near the pyramid. Since the arcs in Figure 2 represent horizontal reflections, any elevation above an arc in the upper portion of the figure (but below the summer solstice limit) would have sent the reflected beam below horizontal to a point on the ground closer to the pyramid. There are also situations in which a reflected beam might be seen cast upward. For example, if there was a layer of clouds over the pyramid at sunrise or sunset, but the sun was low enough to illuminate the pyramid beneath the cloud deck, the specular reflection would have lit up the under surface of the clouds directly above the pyramid. In this situation the tables are turned: the sun is on the horizon and the reflection is toward the sky.

Reflections of bright celestial objects other than the sun would perhaps have been visible at night in the polished sides of the pyramids. The full moon and its gibbous phases would have been especially bright, perhaps luminous enough for its image to be seen from remote sites. With its monthly cycle superimposed on a yearly cycle, the moon would have followed a more complex schedule of reflections than that of the sun. Other sources – bright planets and stars – might have added interesting background to the mirror images of the sky.

Given the central role of the sun in their culture, it is probable that, Egyptians would have considered the solar specular effect in their architecture. Attuned as they were to the natural world, and to the sun in particular, the Egyptians would have attached meaning to any special solar alignments inherent in the configuration of their pyramids. To be sure, pyramid designs would have been driven by more basic requirements of location and orientation, but the builders might have made adjustments to take advantage of specular effects. Unique relationships revealed in the reflections, even after the fact, could be taken as confirmation of divine provenance. Whether part of the original plan or revealed later as a consequence, specular flashes from the Great Pyramid, together with its ever-visible diffuse shine, would have been impressive at any distance, a stunning reminder of the magnificence of the Egyptian world.

## Acknowledgments

The author greatly appreciates conversations with Dr. Janet Jones of Bucknell University that led to this work. The author is also grateful to the Global Monitoring Laboratory of the NOAA Earth System Research Laboratories and Movable Type Scripts for making their interactive calculators available online.

See: E. Nell & C. Ruggles, "The Orientations of the Giza pyramids and associated structures", *Journal for the History of Astronomy*, Vol. 45, No. 3, p.304 (2014).
https://arxiv.org/vc/arxiv/papers/1302/1302.5622v1.pdf

# Appendix 1

## Derivation of the Sun's Position

For each side of the pyramid the required location of the sun is determined by the direction of the observer. Once the horizontal reflection azimuth and the slope of the pyramid face are given, the solar position vector can be calculated. We begin with the east side of the pyramid. Define an x, y, z frame aligned with the axes of the pyramid, with x aligned east, y aligned north and z vertical. Azimuth is measured clockwise looking down on the pyramid (Figure 1). The slope angle of the face, $\phi$, is measured between vertical and the normal to the face (or between the base and the side). To find the position vector of the sun when its light is reflected toward the observer:

1) Create a unit vector in the horizontal x, y plane pointed from the observer toward the pyramid at azimuthal angle $\theta$. This is the direction from which the observer sees the image of the sun in the side of the pyramid:

$O = [x, y, z] = [-\sin\theta, -\cos\theta, 0]$

2) Create a new frame x', y, z' in the plane of the pyramid's east side by rotating about y by the slope angle $\phi$. The vector from the observer in the new frame is:

$O' = [-\sin\theta\cos\phi, -\cos\theta, -\sin\theta\sin\phi]$

3) Change the z' component into –z' to form the reflection of $O'$. The new vector points to the sun:

$S' = [-\sin\theta\cos\phi, -\cos\theta, \sin\theta\sin\phi]$

4) Rotate back to the x, y, z frame. The result is the sun's position vector in the original coordinates. The solar vector $S_E$ for the east side of the pyramid is:

$$S_E = [-\sin\theta\cos2\phi, -\cos\theta, \sin\theta\sin2\phi]$$

5) In a similar manner, create the solar vector for each of the other three sides. With the observer's azimuth $\theta$ measured from north in all cases, the solar position vectors for south, west and north are

$$S_S = [-\sin\theta, -\cos\theta\cos2\phi, -\cos\theta\sin2\phi]$$
$$S_W = [-\sin\theta\cos2\phi, -\cos\theta, -\sin\theta\sin2\phi]$$
$$S_N = [-\sin\theta, -\cos\theta\cos2\phi, \cos\theta\sin2\phi]$$

6) Calculate the sun's azimuth and elevation from the position vectors $S_E, S_S, S_W, S_N$. Using the notation $S = [S_x, S_y, S_z]$, the solar location required to direct the reflected light beam to the observer from each side is:

*Solar azimuth* = $\arctan2(S_y, S_x)$
*Solar elevation* = $\arcsin(S_z/\sqrt{S_x^2 + S_y^2 + S_z^2})$

Here "arctan2" is the arctangent modified to remove the azimuth ambiguity. Solar azimuth can be positive or negative (measured clockwise or counterclockwise). Solar elevation is the angle above horizontal. Only positive elevations (sun above the horizon) can produce reflections. Not all sun positions can be reached in practice because they are restricted to the range between winter and summer solstices at the pyramid latitude (see Figure 2).

The equations developed here can be adapted to any structure that has flat, sloping sides. A rotation away from north-south alignment is modeled by replacing the observer's azimuth $\theta$ with $\theta-\rho$, where $\rho$ is the rotation angle, and then adding $\rho$ to the resultant solar azimuth. With this modification it is possible to model reflections from tetrahedral pyramids and other polyhedral structures with arbitrary slopes and alignments.